\documentclass[a4paper,11pt,twoside]{article}

\usepackage[a4paper,top=3cm, bottom=3cm, left=3cm, right=3cm]{geometry}
\usepackage[american]{babel}
\usepackage[utf8]{inputenc}
\usepackage[T1]{fontenc}
\usepackage{fourier}
\usepackage{bbm}
\usepackage{amssymb,amsmath,amsthm}
\usepackage{latexsym}
\usepackage{authblk}
\usepackage[shortlabels]{enumitem}
\usepackage{stmaryrd}
\usepackage{mathtools}
\usepackage{etoolbox}
\usepackage{needspace}
\usepackage{graphicx,color,xcolor,xparse,easyReview}
\usepackage[colorlinks=true,hyperindex=true]{hyperref}
\colorlet{darkblue}{blue!50!black}
\hypersetup{
	colorlinks,%
	citecolor=darkblue,%
	filecolor=red,%
	linkcolor=darkblue,%
	urlcolor=darkblue,%
	pdfnewwindow=true,%
	pdfstartview={FitH}
}
\newcommand{\tr}{\mathop{\mathrm{tr}}\nolimits}
\newcommand{\e}{\mathrm{e}}
\renewcommand{\d}{\mathrm{d}}

\renewcommand{\i}{\mathrm{i}}
\renewcommand{\atop}[2]{\genfrac{}{}{0pt}{}{#1}{#2}}

\newcommand{\un}{{\mathbb I}}
\newcommand{\ra}{\rightarrow}

\newcommand{\ran}{\mathop{\mathrm{Ran}}}
\renewcommand{\ker}{\mathop{\mathrm{Ker}}\nolimits}
\newcommand{\Ent}{\mathop{\mathrm{Ent}}\nolimits}
\newcommand{\EP}{\mathop{\mathrm{EP}}\nolimits}
\newcommand{\bra}{\langle}
\newcommand{\ket}{\rangle}
\newcommand{\be}{\begin{equation}}
\newcommand{\ee}{\end{equation}}
\newcommand{\bea}{\begin{eqnarray}}
\newcommand{\eea}{\end{eqnarray}}

\newcommand{\cs}{{\cal S}}

\renewcommand{\d}{{\rm d}}

\newcommand{\spec}{\mathop{\mathrm{spec}}\nolimits}
\newcommand{\ad}{\mathop{\mathrm{ad}}\nolimits}

\newcommand{\grintl}{[\kern-.18em [}
\newcommand{\grintr}{]\kern-.18em ]}

\newcounter{resultcounter}[section]

\theoremstyle{plain}
\newtheorem{thm}[resultcounter]{Theorem}
\newtheorem{lem}[resultcounter]{Lemma}
\newtheorem{prop}[resultcounter]{Proposition}
\newtheorem{cor}[resultcounter]{Corollary}
\theoremstyle{definition}
\newtheorem{definition}[resultcounter]{Definition}
\newtheorem{rem}[resultcounter]{Remark}
\def\bed{\begin{definition}}
\def\eed{\end{definition}}

\def\proof{\noindent{\bf Proof.\ }}
 \def\cB{{\cal B}} 
  
 \def\cH{{\cal H}} \def\cI{{\cal I}}
\def\cJ{{\cal J}}  \def\cL{{\cal L}}
\def\cM{{\cal M}}  \def\cO{{\cal O}}
 \def\cQ{{\cal Q}} \def\cR{{\cal R}}
\def\cS{{\cal S}} \def\cT{{\cal T}}

\newcommand{\R}{{\mathbb R}}
\newcommand{\N}{{\mathbb N}}
\newcommand{\Q}{{\mathbb Q}}
\newcommand{\C}{{\mathbb C}}

\newcommand{\E}{{\mathbb E}}

\renewcommand{\P}{{\mathbb P}}

\renewcommand{\i}{{\rm i}}

\def\qed{\hfill $\Box$\medskip}

\newcommand{\qrm}{QRM }
\newcommand{\diag}{{\rm Diag}}

\NewDocumentCommand{\ASS}{mm}{\expandafter\newcommand\csname #1\endcsname{{\hyperref[#1]{\bf (#2)}}}}
\newcommand{\assuming}[3]{
\begin{quote}\label{#2}{\bf(#1) }%
#3%
\end{quote}%
\ASS{#2}{#1}}

\usepackage{fancyhdr}
\title{Two-Time Measurement of Entropy Transfer \\ in Markovian Quantum Dynamics}
\author[1]{Alain Joye}
\author[2]{Claude-Alain Pillet}
\affil[1]{Univ. Grenoble Alpes, CNRS, Institut Fourier, 38000 Grenoble, France}
\affil[2]{Universit\'e de Toulon, CNRS, CPT, UMR 7332, 83957 La Garde, France, \newline
Aix-Marseille Univ, CNRS, CPT, UMR 7332, 13288 Marseille, France}

\begin{document}

\def\today{}
\date{ }

\maketitle

\vspace{1cm}
\begin{small}
\noindent{\bf Abstract.} We consider a protocol for the two-time measurement of
entropic observables in quantum open systems driven out of thermal equilibrium by
coupling to several heat baths. We concentrate on the Markovian approximation of
the time-evolution and relate the expected value of the so defined entropy
variations with the well-known expression of entropy production due to Lebowitz
and Spohn. We do so under the detailed balance condition and, as a byproduct, we
show that the probabilities of outcomes of two-time measurements are given by a
continuous time Markov process determined by the Lindblad generator of the
Markovian quantum dynamics.
\end{small}

\section{Introduction}

In this note, we consider open quantum systems in the Markovian approximation,
and more specifically entropy transfer in such systems out of thermal
equilibrium. We define entropy variation by the two-time measurements of
certain entropic observables and relate the resulting quantities with the
entropy production as defined by Lebowitz and
Spohn~\cite{Spohn1978b,Jaksic2014a}. Entropy production is of prime interest in
nonequilibrium statistical mechanics. Obviously, the two-time measurement
approach to entropy production can also be considered beyond the Markovian
approximation. In the Hamiltonian framework of open quantum systems, where the
joint dynamics of the system and its extended environment is considered, modular
theory, which only surfaces in the Markovian case, provides a rich
mathematical structure. We refer the interested reader
to~\cite{Benoist2023a,Benoist2024b,Benoist2024} for a detailed exposition.
However, given the wide usage of the Markovian approximation in physics, we feel
that a discussion in the latter context is appropriate.

The two-time measurement protocol was first used in the context of quantum
systems out of equilibrium in~\cite{Kurchan2000}, and involves the following
procedure: Initially, say at time $0$, a measurement of a given observable is
performed on the system, resulting in a new state determined by the outcome of
this first measurement. This new state then evolves according to quantum
dynamics up to  some later time $t$, after which a second measurement of the
same observable is performed. This determines the quantum mechanical
probabilities to get an outcome at time $t$, given the outcome at time $0$, and
hence the probabilities of the variations of the observable between times $0$
and $t$.

The definition of entropy production for a Markovian quantum dynamics generated
by a Lindblad operator was motivated by physical considerations on the entropy
balance relation for systems interacting with one or several thermal reservoirs
in~\cite{Spohn1978b}. It was generalized to arbitrary Markovian quantum
dynamics, or quantum dynamical semi-groups, in~\cite{Spohn1978a}. See
also~\cite{Jaksic2014a} for a detailed account on this topic.

Our goal is to relate this approach to the more operational one involving a
two-time measurement of entropic observables. We consider in particular non
equilibrium situations characterized by the fact that the Lindblad generator of
the Markovian quantum dynamics is given by a sum of individual sub-Lindbladians
admitting different invariant states. In all cases, we need these sub-Lindbladians
to satisfy the detailed balance condition. As a byproduct, we remark that the
outcomes of two-time measurements probabilities are given by a continuous time
Markov process on a finite state space determined by the Lindblad  generator of
the Markovian quantum dynamics, under generic hypotheses. This allows us to
express the properties of the quantum measurements in terms of classical data.

Related approaches of the entropy production for different quantum dynamics and
various models have been proposed in the physics literature,
see~\cite{Fiorelli2023} and the references therein for a recent account.

The paper is organized as follows. After setting up Markovian quantum dynamics
in Section~\ref{sectMQD}, Section~\ref{sect2TMP} gathers properties of the
two-time measurement protocols of a quantum observable. We discuss the detailed
balance condition in Section~\ref{secDB} and recall the original Lebowitz-Spohn
definition of entropy production and its main properties in
Section~\ref{sectEP}. We then state and prove our main result regarding the
relation of entropy production and two-time measurement
of the entropy observable in Section~\ref{secMainResults1}.
Then, in Section~\ref{secClassicMarkov}, we turn to
the definition of the classical Markov process related to the two-time
measurement protocol. In Section~\ref{secHeritage}, we invoke this classical
Markov process to investigate some specific properties of the two-time
measurement protocol when the environment of the system is in thermal
equilibrium. We elaborate on the use of the underlying classical Markov process
to express the moments generating function of the latter in terms of classical
properties under certain circumstances in Section~\ref{secMGF}. The paper closes
with Section~\ref{secExample}, where we provide an example, the so-called
quantum reset model.

\section{Markovian description of nonequilibrium open quantum systems}
\label{sectMQD}

We consider a quantum system with finite dimensional Hilbert space $\cH$.
Observables of the system are elements of $\cO$, the $C^\ast$-algebra of all
linear operators on $\cH$, and we denote by $\cB(\cO)$ the set of {\sl
super-operators,} {\it i.e.,} linear operators on $\cO$. Below, $\cT$ denotes the
vector space of linear operators on $\cH$ equipped with the trace norm
$\|A\|_1=\tr(\sqrt{A^\ast A})$. We introduce a duality bracket on
$\cT\times\cO$ by setting
$$
\bra A | B\ket=\tr (A^* B),
$$
and denote the adjoint of a linear map $\cM:\cT\ra\cT$ with respect to this
duality by $\cM^{\dagger}\in\cB(\cO)$. States of our system are described by
density matrices, {\sl i.e.,} elements of the convex set
$$
\cS=\{\rho\in\cT\mid\rho\geq 0, \tr(\rho)=1\}.
$$
A state $\rho\in\cS$ is said to be faithful, written $\rho>0$, whenever $\ker\rho=\{0\}$.

The effective evolution equation of states within the Markovian framework is
$$
\dot{\rho}(t) = \cL (\rho(t)), \quad t \in [0,\infty), \quad \rho(0) = \rho_0 \in \cS,
$$
where the generator is the {\sl Lindbladian}
\be\label{genlib}
\cL(\,\cdot\;)=-\i[H,\,\cdot\;]
+ \sum_l\left(\Gamma_l\cdot \Gamma_l^\ast-\frac12\big\{\Gamma_l^\ast\Gamma_l,\,\cdot\;\big\}\right).
\ee
In the above equation, the sum is over a finite set of indices, $H$ and the
$\Gamma_l$ are all elements of $\cO$, with $H$ being self-adjoint. The
Hamiltonian part of the Lindbladian~\eqref{genlib}, represented by the
commutator with $H$, describes the state's evolution in the absence of an
environment. The dissipator, which is the second term in~\eqref{genlib}, encodes
the global effect of the environment on the evolution.

The family $(\e^{t\cL})_{t\ge0}$ is a norm continuous semi-group of completely
positive trace-preserving (CPTP) contraction on $\cT$, defining a  {\sl Markov
quantum dynamics} (MQD) on $\cS$, see~\cite{Lindblad1976,Gorini1976}. Any
element of $\cS\cap\ker\cL$ represents a {\sl steady state} of this MQD. The set
of such states is never empty. The MQD is called {\sl relaxing} if
$$
\lim_{t\to\infty}\e^{t\cL}(\rho_0)=\rho^+
$$
holds for some $\rho^+\in\ker\cL$ and all $\rho_0\in\cs$. In this case, $\ker\cL$ is the
one-dimensional subspace of $\cT$ spanned by $\rho^+$.
A general algebraic condition on the Kraus operators
$\Gamma_l$ which ensures the relaxing property has been obtained in~\cite[Theorem~2]{Spohn77b}.

Even though the representation~\eqref{genlib} of a Lindbladian can sometimes be
deduced from physics, e.g., through the Davies weak coupling
limit~\cite{Davies1974,Davies1976a}, this representation is in general not
unique. For the Lindbladians occurring in the following, we tacitly assume that
such a representation has been chosen.

\medskip
In this note we will assume, without further mention, that the environment of
the system of interest consists of several reservoirs, inducing the
following structure of the Lindbladian $\cL$.

\assuming{RS}{RS}{(1) The Lindbladian $\cL$ can be decomposed as
\be\label{decomplind}
\cL = \sum_{j\in\cJ}\cL_j,
\ee
each {\sl sub-Lindbladian} $\cL_j$ generating the MQD of the system coupled to a reservoir $\cR_j$.

(2) Each MQD $(\e^{t\cL_j})_{t\ge0}$ has a unique faithful steady state denoted by $\rho_j^+$.
}

\section{Two-Time Measurement Protocol}\label{sect2TMP}

Let us consider here the two-time measurement protocol (2TMP for short) of a
quantum observable $S=S^\ast\in\cO$, for states that vary in time according to the
MQD $(\e^{t\cL})_{t\geq 0}$ generated by a Lindbladian $\cL$. This procedure
yields statistical information on the variation of the observable $S$ with
time.

The initial state is $\rho_0\in\cS$ and the observable $S$ admits the spectral
decomposition\footnote{$\spec(A)$ denotes the spectrum of $A\in\cO$.}
$$
S=\sum_{s\in\spec(S)} s P_s.
$$
A measurement of $S$ in the state $\rho_0$ has the outcome $s\in\spec(S)$ with
probability\footnote{In the following, the subscript to the observable $S$
refers to the time at which the measurement is performed.}
\be\label{iniprob}
\P_{\rho_0}(S_0=s)=\langle\rho_0 |P_s\rangle=\tr (P_s\rho_0P_s),
\ee
and, according to the reduction postulate, the state undergoes the transformation\footnote{For our purpose,
outcomes with zero probability are irrelevant.}
$$
\rho_0 \mapsto \frac{P_s\rho_0 P_s}{\tr (P_s\rho_0P_s)}
$$
immediately after the measurement. After evolving this state for a time $t>0$, a
second measurement of the observable $S$ has the outcome $s'\in\spec(S)$ with
probability
\be\label{kifj}
\P_{\rho_0}(S_t=s'|S_0=s)=\frac{\langle\e^{t\cL}(P_s\rho_0 P_s)|P_{s'}\rangle}{\tr (P_s \rho_0 P_s)}.
\ee
According to Bayes rule, the joint law for the outcome of the two-time
measurement is
\be\label{jandk}
\P_{\rho_0}(S_0=s\, \&\, S_t=s')= \langle\e^{t\cL}(P_s\rho_0 P_s)|P_{s'}\rangle.
\ee
Let us note here that this formula leads to the following expression for the law
of the outcomes of the second measurement
\be\label{almtopro}
\P_{\rho_0}(S_t=s')=\sum_{s\in\spec(S)}\langle\e^{t\cL}(P_s\rho_0 P_s)|P_{s'}\rangle=\langle\e^{t\cL}(\diag_S (\rho_0))|P_{s'}\rangle,
\ee
where $\diag_S$ denote the CPTP map defined on $\cT$ by
$$
T\mapsto\diag_S(T)=\sum_s{P_sT P_s}.
$$
\begin{rem}\label{remDiag}
We will denote by the same symbol the linear map on $\cO$ defined by the same
formula. With this convention, $\diag_S$ is the self-adjoint
($\diag_S^\dagger=\diag_S$) projection onto the commutant
$\{S\}'=\{T\in\cT\mid[T,S]=0\}$ of $S$. For later reference, we observe that
$\diag_S$ actually only depends on the spectral projections of $S$ so that for
any injection $F:\spec(S) \mapsto \R$, we have $\diag_S=\diag_{F(S)}$.
\end{rem}

Note also the {\sl decoherence effect} of the first measurement: If
$[\rho_0,S]\neq0$ and $t>0$, the right-hand side of~\eqref{almtopro} is in general
different from $\langle\e^{t\cL}(\rho_0)|P_{s'}\rangle$, which is the
probability for a measurement of $S$ performed at time $t$ to have the outcome
$s'$ if the system was started at time $0$ in the state $\rho_0$, {\sl without}
measuring $S$ at time zero.

From Relation~\eqref{jandk}, we immediately derive the law $\Q_{\rho_0}^t$ of
the variation of $S$ during the time interval $[0,t]$ according to the two-time
measurement of $S$ in the state $\rho_0$
\be\label{2tmp}
\Q ^t_{\rho_0}(\Delta S=\sigma)
=\sum_{\atop{s,s'\in\spec(S)}{s'-s=\sigma}}\langle\e^{t\cL}(P_s\rho_0 P_s)|P_{s'}\rangle.
\ee
The moment generating function of the random variable $\Delta S$ thus defined is given by
$$
e^t_{\rho_0}(\alpha)=\E^{t}_{\rho_0}(\e^{\alpha\Delta S})
=\sum_{s,s'\in\spec(S)}\e^{\alpha(s'-s)}\langle\e^{t\cL}(P_s\rho_0 P_s)|P_{s'}\rangle.
$$
Performing the summation over $s'$ yields
$$
e^t_{\rho_0}(\alpha)=\sum_{s\in\spec(S)}\e^{-\alpha s}\langle\e^{t\cL}(P_s\rho_0 P_s)|\e^{\alpha S}\rangle,
$$
and hence
$$
e^t_{\rho_0}(\alpha)
=\sum_{s\in\spec(S)}\langle\e^{t\cL}(P_s\rho_0 P_s\e^{-\alpha S})|\e^{\alpha S}\rangle
=\langle\e^{t\cL}(\diag_S(\rho_0)\e^{-\alpha S})|\e^{\alpha S}\rangle.
$$
The expected value of $\Delta S$ is given by\footnote{$\un$ denotes the unit of $\cO$.}
\begin{align}
\E^{t}_{\rho_0}(\Delta S)=(\partial_\alpha e^t_{\rho_0})(0)
&=\langle\e^{t\cL}(\diag_S(\rho_0))|S\rangle
-\langle\e^{t\cL}(\diag_S(\rho_0)S)|\un\rangle\nonumber\\[8pt]
&=\langle\e^{t\cL}(\diag_S(\rho_0))|S\rangle-\langle\rho_0|S\rangle,
\label{eq:expgen}
\end{align}
where we used the fact that the dual semi-group $\e^{t\cL^\dagger}$ is
unit-preserving, and the cyclicity of the trace. Finally, the expected
value of $\Delta S$
writes as a difference of quantum mechanical expectation values of the
observable $S$ in states at time $t$ and time $0$
where the initial state is indeed $\rho_0$, but the state at time $t$ has been
affected by the decoherence induced by the first measurement.

We note for later reference that
\be
e_{\rho_0}^t(\alpha)=\langle\e^{t\cL_\alpha}(\diag_S(\rho_0))|\un\rangle,
\label{eq:LalphaForm}
\ee
where $\cL_\alpha(\,\cdot\;)=\cL(\,\cdot\;\e^{-\alpha S})\e^{\alpha S}$.
This should be compared with~\cite[Definition~(6)]{Jaksic2014a}, which
is not obviously related to two-time measurements.

\begin{rem}Davies weak coupling limit~\cite{Davies1974,Davies1976a} provides a relation between this two-time measurement
of the system observable $S$ and idealized two-time measurements of reservoir energies
in the Hamiltonian description of the open system, patterned on the protocol
introduced in~\cite{Kurchan2000}, the latter being itself related to
the full counting statistics of charge transport introduced in the physics
literature~\cite{Levitov1996}, see also~\cite{Derezinski2008} and~\cite[Section~4.2]{Jaksic2010b}.
We refer the reader to~\cite[Section~5]{Jaksic2014a}
for a discussion of this relation, and to~\cite{Benoist2024b} for a justification
of the involved idealization by appropriate thermodynamic limits of the reservoirs.
\end{rem}

\medskip
Further, assuming the MQD to be relaxing to the steady state $\rho^+$,
we get the following large $t$ limits of the above quantities
\be\label{eq:limcopro}
\begin{split}
\lim_{t\to\infty}\langle\e^{t\cL}(P_s\rho_0 P_s)|P_{s'}\rangle
 &=\langle\rho^+|P_{s'}\rangle \langle\rho_0|P_s\rangle,\\[8pt]
\lim_{t\to\infty}\P_{\rho_0}(S_t=s'|S_0=s)
 &=\langle\rho^+|P_{s'}\rangle,\\[8pt]
\lim_{t\to\infty}\Q^t_{\rho_0}(\Delta S=\delta)
 &=\sum_{\atop{s,s'\in\spec(S)}{s'-s=\delta}}\langle\rho^+|P_{s'}\rangle\langle\rho_0|P_s\rangle,\\[8pt]
 \lim_{t\to\infty}e^t_{\rho_0}(\alpha)
  &=\langle\rho^+|\e^{\alpha S}\rangle\langle\rho_0|\e^{-\alpha S}\rangle,\\[8pt]
\lim_{t\to\infty}\E^t_{\rho_0}(\Delta S)
 &=\langle\rho^+-\rho_0|S\rangle.
\end{split}
\ee
The RHS of the last relation is the difference of the QM expectation
of $S$ in the limiting state at time $t=\infty $, $\rho^+$, and in
the initial state $\rho_0$.

\section{Detailed Balance}\label{secDB}

Let us briefly recall the notion of detailed balance (DB for short)
in the Lindblad context, we refer the reader to~\cite{Agarwal1973,Alicki1976,Fagnola2007,Fagnola2010} for details.

The dual $\cL^\dagger$ of the general Lindblad operator~\eqref{genlib} takes
the form
\be\label{ldagphi}
\cL^\dagger(\,\cdot\;)=\i[H,\,\cdot\;]-\frac12\big\{\Phi(\un),\,\cdot\;\big\}+\Phi(\,\cdot\;),
\ee
where $\Phi(\,\cdot\;)=\sum_l\Gamma_l^\ast\cdot\Gamma_l\in\cB(\cO)$ is a Completely Positive (CP) map.
With a slight abuse of language, we will say that $\Phi$ is the CP-map associated to  $\cL$.

For a faithful state $\rho>0$, we introduce the following $\rho$-inner product on $\cO$,
$$
\bra A |B\ket_\rho:=\tr(\rho A^\ast B).
$$
We denote by $\cM^\rho$ the adjoint of a super-operator $\cM\in\cB(\cO)$ w.r.t.\;this
inner product, that is
$$
\bra A| \cM(B)\ket_\rho=\bra \cM^\rho(A)|B\ket_\rho,
$$
for all $A,B\in\cO$.This is easily seen to be equivalent to
\be\label{symrho}
\cM^\rho(A)=\cM^\dagger(A\rho)\rho^{-1}.
\ee
We will say that $\cM$ is $\rho$-self-adjoint whenever $\cM^\rho=\cM$.

\medskip
Consider a pair $(\rho,\cL)$ where $\cL$ is a Lindbladian and $\rho\in\cs\cap\ker\cL$.
The following condition essentially characterizes an open system in contact with
a reservoir in thermal equilibrium (see~\cite{Kossakowski1977}).

\assuming{DB}{DB}{$(\rho,\cL)$ satisfies the detailed balance condition if
$\rho>0$ and the CP-map $\Phi$ associated to $\cL$ is $\rho$-self-adjoint.
}

In the context of~\RS, the next assumption further specifies the inverse temperature
$\beta_j>0$ of reservoir $\cR_j$.

\assuming{KMS}{KMS}{For each pair $(\cL_j,\rho_j^+)$ of Assumption~\RS,
one has\footnote{$F_j=-\beta_j^{-1}\log\,\tr(\e^{-\beta_j H})$ is the free
energy of the system at inverse temperature $\beta_j$.}
$$
\cL^\dagger_j(\,\cdot\;)=\i[H,\,\cdot\;]-\frac12\big\{\Phi_j(\un),\,\cdot\;\big\}+\Phi_j(\,\cdot\;),\qquad
\rho_j^+=\e^{-\beta_j(H-F_j)}.
$$
}

For our purpose, the following simple consequence of the detailed balance condition will be important:
\begin{lem}\label{comdl}
Assume \DB{} holds for $(\rho,\cL)$, then,
\be\label{comrhol}
\diag_\rho \circ \cL = \cL \circ \diag_\rho.
\ee
\end{lem}

\begin{rem}
\begin{enumerate}[label=(\roman*)]
\item By Remark~\ref{remDiag}, the statement is equivalent to
$\diag_\rho \circ \cL^\dagger = \cL^\dagger \circ \diag_\rho$.
\item As the proof shows, the dissipator part of $\cL^\dagger(\cdot)$, $-\frac{1}{2}\big\{\Phi(\un),\,\cdot\;\big\}+\Phi(\,\cdot\;)$ is $\rho$-self-adjoint under Assumption~\DB.
\end{enumerate}
\label{remsal}
\end{rem}
\proof
Let $\Phi$ be the CP-map associated to $\cL$ and define $\Delta_\rho\in\cB(\cO)$
by\footnote{$\Delta_\rho$ is the modular operator of $\rho$.}
$$
\Delta_\rho(X)=\rho X \rho^{-1}.
$$
The cyclicity of the trace implies that that $\Delta_\rho^\dagger=\Delta_\rho$.
Moreover, under~\DB, see~\cite[Theorem~7.1]{Jaksic2014a},
\be\label{comodul}
[\Phi,\Delta_\rho]=0.
\ee
This relation is the root of the following identities
\be\label{comuseful}
\ [\Phi(\un),\rho]=0, \qquad [H,\rho]=0.
\ee
Indeed, applied to the unit $\un$, \eqref{comodul} immediately yields the first identity.
It follows that
$$
0=\cL(\rho)= -\i[H,\rho] -\Phi(\un)\rho+\Phi^\dagger(\rho),
$$
and since, by Relation~\eqref{symrho} and~\DB,
$\Phi^\dagger(\rho)=\Phi^\rho(\un)\rho=\Phi(\un)\rho$, the
second identity in~\eqref{comuseful} follows.

From~\eqref{comuseful}, it easily follows that the first two terms in~\eqref{ldagphi}
commute with $\diag_\rho$. It remains to show that $\Phi$ commutes with
$\diag_\rho$ as well. To this end, we first deduce from~\eqref{comodul}
that $\Phi$ commutes with the spectral projections of $\Delta_\rho$.
Now clearly $1\in\spec(\Delta_\rho)$ with $\ker(\Delta_\rho-\mathrm{Id})=\{\rho\}'$.
It follows from Remark~\ref{remDiag} that $\diag_\rho$ is the spectral projection
of $\Delta_\rho$ to its eigenvalue $1$ which ends the proof of~\eqref{comrhol}.

Finally, the properties $\Phi=\Phi^\rho$ together with~\eqref{comuseful} show
that under~\DB, the dissipator of $\cL^\dagger$ is $\rho$-self-adjoint, which
justifies Remark~\ref{remsal}-(ii).
\qed

\begin{rem}
More generally, one can consider the inner products on $\cO$ defined by
$\bra A|B\ket_{\rho_s}:=\tr(\rho^sA^*\rho^{1-s}B)$, where $s\in[0,1]$.
A pair  $(\rho,\cL)$ with $\ker\cL\ni\rho>0$ satisfies the
$\rho_s$-Detailed Balance Condition whenever the associated CP map $\Phi$
is self-adjoint w.r.t.\;this inner product. It appears that the conclusion
of Lemma~\ref{comdl} holds true if the pair $(\rho,\cL)$ satisfies the
$\rho_s$-Detailed Balance Condition for some $s\in [0,1]\setminus\{1/2\}$.
Indeed, \cite[Lemma~2.5]{Carlen2017} ensures $[\Phi,\Delta_\rho]=0$, from which the
first identity in~\eqref{comuseful} follows. One uses the expression
$$
\Phi(\,\cdot\;)=\Phi^{\rho_s}(\,\cdot\;)
=\rho^{s-1}\Phi^\dagger(\rho^{1-s}\cdot\rho^s)\rho^{-s}
$$
acting on the unit $\un$ to deduce
$$
\Phi^\dagger(\rho)=\rho^{1-s}\Phi(\un)\rho^s=\rho\Delta_\rho^{-s}(\Phi(\un))
=\rho\Phi(\Delta_\rho^{-s}(\un))=\rho\Phi(\un),
$$
which implies $[H,\rho]=0$ as above. The rest of the proof is identical.
Thus, for $s\in [0,1]\setminus\{1/2\}$ the various notions of detailed balance
are all equivalent for our purposes (see also~\cite[Section~8]{Fagnola2007}).

This is not true for the special case $s=1/2$, which corresponds to the
so called {\sl KMS detailed balance condition,} as the following counter example
stemming from~\cite[Example~38]{Fagnola2007} shows (see also~\cite[Appendix~B]{Carlen2017} and~\cite[Appendix~B]{Benoist2022}).

The Hilbert space is $\cH=\C^2$ and the dual Lindbladian
$$
\cL^{\dagger}(X)=\i[H,X]-\frac12\{\Phi(\un),X\}+\Phi (X)
$$
acts on $2\times2$ matrices. With $\sigma_{j}$, $j\in\{1,2,3\}$ the Pauli matrices,
and $\sigma_0=\un$, the example is defined by
\[
H=\kappa\omega\sigma_1,\quad
\Gamma=\sqrt{1-\kappa^2}\sigma_0+\i r\sigma_1+s\sigma_2+\sigma_3,\quad
\Phi(X)=\Gamma^{\ast}X\Gamma,\quad
\Phi^\dagger(\rho)=\Gamma\rho\Gamma^{\ast},
\]
where, $\kappa\in(0,1)$, $\omega,s,r\in\R$. The relation with the notation
in~\cite{Fagnola2007} is $\kappa=2\sqrt{\nu(1-\nu)}$ and $\kappa\omega=\Omega$,
with the restriction $\nu\in(0,1/2)$ which ensures that $1-2\nu=\sqrt{1-\kappa^2}$.
Setting
\[
s=\omega\frac{1+\kappa}{1-\kappa},\qquad
r=\omega\sqrt{\frac{1+\kappa}{1-\kappa}},
\]
ensures that
\[
\rho=\frac12\left(\sigma_0-\sqrt{1-\kappa^2}\sigma_3\right)
=\left(
   \begin{array}{cc}
     \nu & 0\\
      0 & 1 - \nu
   \end{array}
  \right)
\]
is invariant
\[
\cL (\rho) = 0,
\]
and that
\[
\Delta^{1/2}_{\rho} \Gamma^{\ast}=\rho^{1/2}\Gamma^\ast\rho^{-1/2}=\Gamma.
\]
The latter implies that $\Phi$ is $\rho_{1/2}$-self-adjoint
\[
\Phi^{\rho_{1/2}}=\Phi
\]
so that the KMS-detailed balance condition is satisfied, but one finds that
\[
\i[H,\rho]=-\kappa\sqrt{1-\kappa^2}\sigma_1\neq 0, \quad \text{and}\quad
(\cL\circ\diag_{\rho}-\diag_{\rho}\circ\cL)(\sigma_1)=-4\omega\frac\kappa{1-\kappa}\sigma_3\neq0.
\]
\end{rem}

\section{Entropy production of MQD}\label{sectEP}

The relative entropy of a state $\mu\in\cS$ w.r.t.\;another state $\nu$ is defined by
the expression
\begin{align*}
\Ent(\mu|\nu)=
\begin{cases}
  \langle\mu|\log(\mu)-\log(\nu)\rangle, & \text{if } \ker (\nu)\subset \ker (\mu);\\
  +\infty, & \text{otherwise,}
\end{cases}
\end{align*}
which is the immediate extension of the relative entropy of two probability
measures to the non-commutative setting of quantum mechanics. It satisfies
$\Ent(\mu|\nu)\ge0$, with equality iff $\mu=\nu$, as well as Uhlmann's
monotonicity theorem
$$
\Ent(\phi(\mu)|\phi(\nu))\le\Ent(\mu|\nu)
$$
for any CPTP map $\phi$.

The entropy production (EP for short\footnote{It should more accurately be called entropy production {\sl rate.}})
of the MQD $(\e^{t\cL})_{t\geq 0}$ in the state $\rho$ was defined in~\cite[Definition~1]{Spohn1978a}, \cite[Theorem~2]{Spohn1978b}
as
\be\label{entprodgen}
\EP(\rho)=\sum_{j\in\cJ}\EP_j(\rho),\qquad
\EP_j(\rho)=\left.-\frac{\d\ }{\d t}\Ent(\e^{t\cL_j}(\rho)|\rho_j^+)\right|_{t=0},
\ee
where each $\EP_j(\rho)$ represents the entropy production due to the interaction
of the system with the $j^\mathrm{th}$ reservoir. Note that
since $\rho_j^+=\e^{t\cL_j}(\rho_j^+)$, it follows from Uhlmann's theorem
that the function
$$
t\mapsto\Ent(\e^{t\cL_j}(\rho)|\rho_j^+)
$$
is monotone decreasing, so that $\EP_j(\rho)\ge0$.
As proven in~\cite[Theorem~3(i)]{Spohn1978a}, \cite[Theorem~2(ii)]{Spohn1978b}, the map
$$
\cS\ni\rho\mapsto\EP_j(\rho)\in[0,\infty]
$$
is convex. It is given by
\begin{align}\label{entprodfor}
\EP_j(\rho)=\langle\cL_j(\rho)|\log(\rho_j^+)-\log(\rho)\rangle,
\end{align}
where the first term $\langle\cL_j(\rho)|\log(\rho_j^+)\rangle$ is finite since $\rho_j^+>0$. Whenever $0\in\spec(\rho)$,
the second term should be computed in the eigenbasis $\{\phi_r\}$ of $\rho$,
$$
\langle\cL_j(\rho)|\log(\rho)\rangle=\sum_r (\phi_r,\cL(\rho)\phi_r)\log(r)
$$
with the convention\footnote{Note that
whenever $\phi\in\ker\rho$, one must have $(\phi,\cL(\rho)\phi)\ge0$.}
$$
a\log0=\begin{cases}
0&\text{if } a=0;\\[4pt]
-\infty&\text{otherwise.}
\end{cases}
$$

Considering the von~Neumann entropy
$$
S(\rho) = -\langle\rho|\log(\rho)\rangle\geq 0,
$$
of a state $\rho\in\cS$, we have
$$
\frac{\d\ }{\d t}S(\e^{t\cL}(\rho))=-\langle\cL(\e^{t\cL}(\rho))|\log(\e^{t\cL}(\rho))\rangle
=-\sum_{j\in\cJ}\langle\cL_j(\e^{t\cL}(\rho))|\log(\e^{t\cL}(\rho))\rangle
$$
with the same convention. It immediately follows from~\eqref{entprodfor} that
\be\label{eq:Ebalance}
\frac{\d\ }{\d t}S(\e^{t\cL}(\rho))
=\EP(\e^{t\cL}(\rho))-\sum_{j\in\cJ}\langle\cL_j(\e^{t\cL}(\rho))|\log(\rho_j^+)\rangle,
\ee
To interpret this relation, we express the $j^\text{th}$-term
of the sum in its right-hand side using the dual maps $\cL_j^\dagger\in\cB(\cO)$ and the observable
\be\label{eq:SjDef}
S_j^+=-\log(\rho_j^+)
\ee
representing the entropy of the $j^\text{th}$-reservoir. By duality,
$$
-\langle\cL_j(\e^{t\cL}(\rho))|\log(\rho_j^+)\rangle
=\langle\cL_j(\e^{t\cL}(\rho))|S_j^+\rangle
=\langle\e^{t\cL}(\rho)| \cL_j^\dagger(S_j^+))\rangle
=\langle\e^{t\cL}(\rho)|\cI_j^+\rangle,
$$
where $\cI_j^+=\cL_j^\dagger(S_j^+)=\partial_t\e^{t\cL_j^\dagger}(S_j^+)|_{t=0}$ is the
entropy flux observable out of this reservoir. Thus, the identity~\eqref{eq:Ebalance}
becomes the entropy balance relation
\be
\frac{\d\ }{\d t}S(\e^{t\cL}(\rho))
=\EP(\e^{t\cL}(\rho))+\sum_{j\in\cJ}\langle\e^{t\cL}(\rho)|\cI_j^+\rangle,
\label{eq:EntropyBalance}
\ee
which expresses the rate of change in the entropy of the system
as the sum of the entropy production rate and the total entropy flux out of the reservoirs.

In the special case of a steady state $\rho^+\in\cS\cap\ker\cL$, the entropy is constant and the entropy
balance reduces to
\be\label{entsteady}
\EP(\rho^+)=-\sum_{j\in\cJ}\langle\rho^+|\cI_j^+\rangle.
\ee

Finally, in case the decomposition~\eqref{decomplind} is trivial, Formula~\eqref{entprodfor}
becomes
\be\label{singlep}
0\leq \EP(\rho)=\langle\cL(\rho)|\log(\rho^+)-\log(\rho)\rangle.
\ee
The entropy observable is $S^+=-\log(\rho^+)$, the entropy flux observable is
$\cI^+=\cL^\dagger(S^+)$. Note that $\EP(\rho^+)=0=\langle\rho^+| \cI^+\rangle$ in this case.

\medskip

As noted above, it may happen that the time derivative of $S(\e^{t\cL}(\rho))$ and hence $\EP(\e^{t\cL}(\rho))$ become infinite, even for faithful $\rho$. The following proposition, proven in Section~\ref{sec:proof5.1},
gives sufficient conditions that exclude this behavior.
\begin{prop}\label{prop:hebo}
Assume the Lindbladian $\cL$ to be relaxing to a faithful state $\rho^+$.
\begin{enumerate}[label=(\roman*)]
\item For all faithful $\rho\in\cS$ and all $t\ge0$, $\partial_tS(\e^{t\cL}(\rho))<\infty$.
\item For all $\rho\in\cS$ there exists $T(\rho)>0$ such that $\partial_tS(\e^{t\cL}(\rho))<\infty$ for all $t>T(\rho)$.
\end{enumerate}
\end{prop}

\section{Link between EP and 2TMP}
\label{secMainResults1}

We can now establish the link between the Lebowitz-Spohn EP defined
by~\eqref{entprodgen} and the 2TMP of the entropic observable~\eqref{eq:SjDef}, under DB
conditions on the sub-Lindbladians $\cL_j$. We start with the simplest case
where the decomposition~\eqref{decomplind} is trivial, {\sl i.e.,} there is only
one reservoir.
\begin{lem}\label{expflu}
Suppose that~\DB{} holds for the pair $(\rho^+,\cL)$ and set $S^+=-\log\rho^+$.
For any $\rho_0\in\cS$, one has
$$
\E^t_{\rho_0}(\Delta S^+)=\bra\e^{t\cL}(\rho_0)|S^+\ket-\bra\rho_0|S^+\ket.
$$
Thus, there is no decoherence effect of the first measurement and the
expected 2TMP is the difference of the QM expectation value of $S^+$ in the state
$\e^{t\cL}(\rho_0)$ at time $t$ and in the initial state $\rho_0$. In particular
\be
\E^t_{\rho_0}(\Delta S^+)=\int_0^t\bra\e^{s\cL}(\rho_0)|\cI^+\ket\d s
=S(\e^{t\cL}(\rho_0))-S(\rho_0)-\int_0^t\EP(\e^{s\cL}(\rho_0))\d s,
\label{eq:justone}
\ee
where $\cI^+=\cL^\dagger(S^+)$.
\end{lem}

\proof
The expression~\eqref{eq:expgen} and Lemma~\ref{comdl} imply
\be\label{exp2TDB}
\begin{split}
\E^t_{\rho_0}(\Delta S^+)
&=\bra\e^{t\cL}(\diag_{S^+}(\rho_0))|S^+\ket-\bra\rho_0|S^+\ket\\
&=\bra\diag_{S^+}(\e^{t\cL}(\rho_0))|S^+\ket-\bra\rho_0|S^+\ket\\
&=\bra\e^{t\cL}(\rho_0)|\diag_{S^+}(S^+)\ket-\bra\rho_0|S^+\ket\\
&=\bra\e^{t\cL}(\rho_0)|S^+\ket-\bra\rho_0|S^+\ket.
\end{split}
\ee
Applying the fundamental theorem of calculus to the right-hand side
of the last identity, we get
$$
\E^t_{\rho_0}(\Delta S^+)
=\int_0^t\frac{\d\ }{\d s}\bra\e^{s\cL}(\rho_0)|S^+\ket\d s
=\int_0^t\bra\cL(\e^{s\cL}(\rho_0))|S^+\ket\d s
=\int_0^t\bra\e^{s\cL}(\rho_0)|\cL^\dagger(S^+)\ket\d s.
$$
The last equality in~\eqref{eq:justone} now follows from the entropy balance
relation~\eqref{eq:EntropyBalance}.\qed

\begin{rem}
\begin{enumerate}[label=(\roman*)]
\item By differentiation, \eqref{eq:justone} becomes
$$
\frac{\d\ }{\d t}\E_{\rho_0}^t(\Delta S^+)
=-\EP(\e^{t\cL}(\rho_0))+\frac{\d\ }{\d t}S(\e^{t\cL}(\rho_0)).
$$
\item Further, assuming $\lim_{t\to\infty}\e^{t\cL}(\rho_0)=\rho^+$, we have $\int_0^\infty\EP(\e^{s\cL}(\rho_0))\d s=\Ent(\rho_0|\rho^+)$ and
$$
\lim_{t\to\infty}\E^t_{\rho_0}(\Delta S^+)
=S(\rho^+)-S(\rho_0)-\int_0^\infty\EP(\e^{s\cL}(\rho_0))\d s=\bra\rho^+-\rho_0|S^+\ket.
$$
\end{enumerate}
\end{rem}

Thanks to $\EP(\rho)\geq0$, we deduce immediately the following upper bounds on
the 2TMP expectation of the entropic variable $\Delta S^+$ and its time
derivative in terms of the variation of the von~Neumann entropy and its
derivative.

\begin{cor} Under the assumptions of Lemma~\ref{expflu}, the 2TMP expectation of the entropic observable
$S^+=-\log\rho^+$ in the state $\rho_0$ satisfies
$$
S(\e^{t\cL}(\rho))-S(\rho)\ge\E^{t}_{\rho_0}(\Delta S^+).
$$
\end{cor}

\medskip
Turning to the general multi-reservoir case, we apply the two-time measurement protocol to
the MQD $(\e^{t\cL_j})_{t\ge0}$ introduced by~\RS{} and to the entropy observables $S_j^+$
defined in~\eqref{eq:SjDef}.  In this context, for $j\in\cJ$, we set
\be\label{2tmep}
\Q _{j,\rho_0}^t(\Delta S_j^+=\sigma)
=\sum_{\atop{s,s'\in\spec(S_j)}{s'-s=\sigma}}\langle\e^{t\cL_j}(P_{j,s}\rho_0 P_{j,s})|P_{j,s'}\rangle,\qquad
S_j^+=\sum_{s\in\spec(S_j^+)}sP_{j,s}
\ee
and call the random variable $\Delta S_j^+$ the {\sl two-time measurement entropy production}
of the $j^\text{th}$-reservoir.

\begin{thm}\label{ept=0}
Suppose that the Lindbladian $\cL$ satisfies Assumption~\RS{} and is such that,
for all $j\in\cJ$, the pair $(\rho_j^+,\cL_j)$ satisfies~\DB{}. Considering the entropic
observables $S_j^+=-\log(\rho_j^+)$ and $\cI_j^+=\cL_j^\dagger(S_j^+)$, for any $j\in\cJ$, $\rho_0\in\cS$ and $t\ge0$,
one has
\be\label{epvs2tmp}
\E_{j,\rho_0}^t(\Delta S_j^+)
=\int_0^t\bra\e^{s\cL_j}(\rho_0)|\cI_j^+\ket\d s
=S(\e^{t\cL_j}(\rho_0))-S(\rho_0)-\int_0^t\EP_j(\e^{s\cL_j}(\rho_0))\d s,
\ee
where $\E_{j,\rho_0}^t$ denotes the expectation w.r.t.\;the law~\eqref{2tmep}. In
particular, if $\rho^+\in\ker\cL$, then
\be\label{2tmpet}
\EP(\rho^+)=-\lim_{t\downarrow0}
\sum_{j\in\cI}\E_{j,\rho^+}^t\left(\tfrac{\Delta S_j^+}t\right).
\ee
\end{thm}

\proof
Formula~\eqref{epvs2tmp} follows from Lemma~\ref{expflu}, and more
precisely Relation~\eqref{eq:justone}, applied to the pair $(\rho_j^+,\cL_j)$.
Differentiating the first equality in~\eqref{epvs2tmp} gives
$$
\left.\frac{\d\ }{\d t}\sum_{j\in\cJ}\E_{j,\rho^+}^t(\Delta S_j^+)\right|_{t=0}
=\sum_{j\in\cJ}\bra\rho^+|\cI_j^+\ket,
$$
and~\eqref{2tmpet} follows from~\eqref{entsteady}.
\qed

\begin{rem}
\begin{enumerate}[label=(\roman*)]
\item The differential version of~\eqref{epvs2tmp} is
$$
\frac{\d\ }{\d t}\E_{j,\rho_0}^t(\Delta S_j^+)
=\frac{\d\ }{\d t}S(\e^{t\cL_j}(\rho_0))-\EP_j(\e^{t\cL_j}(\rho_0)).
$$
\item For $j\in\cJ$, one has
$$
\EP_j(\e^{t\cL_j}(\rho_0))=-\frac{\d\ }{\d t}\Ent(\e^{t\cL_j}(\rho_0)|\rho_j^+),
$$
so that
$$
\sum_{j\in\cJ}\int_0^\infty \EP_j(\e^{s\cL_j}(\rho_0))\d s=\sum_{j\in\cJ}\Ent(\rho_0|\rho_j^+)
$$
is finite, as opposed to the total entropy production
$$
\sum_{j\in\cJ}\int_0^t \EP_j(\e^{s\cL}(\rho_0))\d s=t\EP(\rho^+)+o(t),\quad(t\to\infty)
$$
which diverges if $\EP(\rho^+)>0$.
\item We note that also
$$
\lim_{t\to\infty}\sum_{j\in \cJ}\E_{j,\rho_0}^t(\Delta S_j^+)
=\sum_{j\in\cJ}\bra\rho_j^+-\rho_0|S_j^+\ket
$$
is finite
\item Under the additional Assumption~\KMS, one has $S_j^+=\beta_j(H-F_j)$ and hence
$\cI_j^+=\beta_j\cQ_j^+$, where $\cQ_j^+=\cL_j^\dagger(H)$ is the observable
describing the heat current out of the $j^\mathrm{th}$ reservoir. We then have
$$
\E_{j,\rho_0}^t(\Delta S_j^+)=\beta_j\int_0^t\bra\e^{s\cL_j}(\rho_0)|\cQ_j^+\ket\d s.
$$
\end{enumerate}
\end{rem}

\section{A Classical Markov Chain}\label{secClassicMarkov}

In this section, we further elaborate on the 2TMP under the DB condition,
showing that a classical continuous time Markov chain can be naturally
associated to the two-time measurement process.

We first recall well known facts about homogeneous continuous time Markov
chains, mainly to set the notation. See, e.g., \cite{Norris1997} for more
details. Such a process $(X_t)_{t\geq 0}$ defined on a finite state space
$\Sigma$ is completely characterized by the initial probabilities of each state $s\in\Sigma$
$$
\P(X_0=s)=\pi_s(0),
$$
and by the probability to find the process in the state $s'\in\Sigma$ at time $t\geq 0$,
given its state $s\in\Sigma$ at time $0$,
$$
\P(X_t=s'|X_0=s)=P_{ss'}(t).
$$
Denoting by $\boldsymbol{1}\in\R^\Sigma$ the column vector whose entries are all set to $1$,
the initial probabilities can be seen as a dual/row vector $\pi(0)=(\pi_s(0))\in\R^{\Sigma\ast}$ normalized by
$\pi(0)\boldsymbol{1}=1$, and  the transition probabilities form a time dependent
matrix acting on $\R^\Sigma$, $P(t)=(P_{ss'}(t))$. This matrix is stochastic, {\sl i.e.,}
such that $P(t)\boldsymbol{1}=\boldsymbol{1}$ for all $t\ge0$, and the family
$(P(t))_{t\ge0}$ is a semi-group
$$
P(t)=\e^{tQ},
$$
generated by the so-called transition rate matrix $Q=(Q_{ss'})$ which
satisfies $Q_{ss'}\geq 0$  for $s\neq s'$ and $Q\boldsymbol{1}=0$.
The time-$t$ probability vector $\pi(t)=(\pi_s(t))$, with $\pi_s(t)=\P(X_t=s)$,
is given by
$$
\pi(t)=\pi(0)P(t).
$$
Hence, an invariant probability vector $\pi^{\text{inv}}$ is characterized by
$\pi^\text{inv}Q=0$. Since $Q\boldsymbol{1}=0$, the Perron-Frobenius theorem ensures the
existence of an invariant probability.

\begin{prop} \label{propmark}
Let $(\rho,\cL)$ satisfy~\DB{} and $\rho_0$ be a faithful initial state. Assume that the entropic observable
$S=-\log\rho$ has simple spectrum, with spectral decomposition $S=\sum_s s P_s$.
Then, there exists a continuous time Markov chain $(X_t)_{t\geq 0}$ on the state
space $\Sigma=\spec(S)$ such that the pair of random variables $(X_0,X_t)$ has the same
statistics as the outcome of the two measurements of $S$ at time $0$ and $t$.
The chain $(X_t)_{t\ge0}$ has the initial probability vector
\be\label{eq:pizero}
\pi_s(0)=\bra\rho_0|P_s\ket,
\ee
and transition matrix
\be\label{eq:cprob}
P_{ss'}(t)=\P_{\rho_0}(S_t=s'|S_0=s)=\bra\e^{t\cL}(P_s)|P_{s'}\ket,
\ee
where $\P^{\rho_0}$ is defined in~\eqref{kifj}.
The corresponding transition rate matrix $Q$ is given by
\be\label{eq:QDef}
Q_{ss'}=\bra P_s|\Phi(P_{s'})\ket-\delta_{ss'}\bra P_{s}|\Phi(\un)\ket
\ee
where $\Phi$ is the CP map associated to $\cL$.
\end{prop}

\begin{rem}
\begin{enumerate}[label=(\roman*)]
\item The conditional probability~\eqref{kifj} takes the $\rho_0$-independent
form~\eqref{eq:cprob} since, for rank one spectral projections,
$P_s\rho_0P_s=P_s\tr(P_s\rho_0P_s)$ for all $s$. In this case, the choice of
initial state $\rho_0$ only manifests itself in the initial probabilities~\eqref{eq:pizero}.
\item Further, assuming the MQD generated by $\cL$ to be relaxing, we deduce from
Relation~\eqref{eq:limcopro} that
$$
\lim_{t\to\infty}\P_{\rho_0}(S_t=s'|S_0=s)=\bra\rho|P_{s'}\ket=\bra\e^{-S}|P_{s'}\ket=\e^{-s'}.
$$
Hence, the invariant probability vector $\pi^\text{inv}$ of the Markov chain
satisfies $\pi_s^\text{inv}=\e^{-s}$. It corresponds to initial states
$\rho_0$  such that $\diag_\rho(\rho_0)=\rho$.
\item  From Definition~\eqref{eq:QDef} we derive, starting from $\e^{-s}P_s=\rho P_s$
and invoking the DB condition $\Phi^\rho=\Phi$,
\begin{align*}
\e^{-s}Q_{ss'}&=\bra\rho P_s|\Phi(P_{s'})\ket-\delta_{ss'}\bra\rho P_{s}|\Phi(\un)\ket\\
&=\bra P_s|\Phi(P_{s'})\ket_\rho-\delta_{ss'}\bra P_{s}|\Phi(\un)\ket_\rho\\
&=\bra\Phi(P_s)|P_{s'}\ket_\rho-\delta_{ss'}\bra P_{s'}|\Phi(\un)\ket_\rho\\
&=\bra\rho P_{s'}|\Phi(P_{s})\ket-\delta_{ss'}\bra\rho P_{s'}|\Phi(\un)\ket=\e^{-s'}Q_{s's},
\end{align*}
which is the classical detailed balance condition for a Markov chain. In matrix form,
it reads
\be\label{mafodb}
RQ=Q^TR,
\ee
where $R_{ss'}=\delta_{ss'}\e^{-s}$. Hence, $P(t)=\e^{tQ}$ has the same property:
\be\label{sympsig}
RP(t)R^{-1}=\e^{tRQR^{-1}}=\e^{tQ^T}=P(t)^T.
\ee
Note that~\eqref{mafodb} implies that $Q$ is self-adjoint with respect to the inner product on $\R^\Sigma$
defined by $R$, so that, in particular, $\spec(Q)\subset (-\infty,0]$.
\item Since under the DB condition, $[\rho,H]=0$ and by assumption the spectrum
of $\rho$ is simple, the transition matrix~\eqref{eq:cprob} is identical to that
of the two-time measurement protocol of the energy observable, in case the
spectrum of $H$ is simple as well.
\end{enumerate}
\label{rem:Markov}
\end{rem}

\proof
The first statement~\eqref{eq:pizero} reformulates Relation~\eqref{iniprob}.
Considering~\eqref{kifj} and Remark~\ref{rem:Markov}-(i), it remains to show
that the RHS of Relation~\eqref{eq:cprob}, {\sl i.e.,} Formula
$P_{ss'}(t)=\bra\e^{t\cL}(P_s)|P_{s'}\ket$ defines a semi-group $(P(t))_{t\ge0}$
of stochastic matrices generated by a matrix $Q$ satisfying Relation~\eqref{eq:QDef}. We first
remark that the stochasticity follows directly from the first equality in~\eqref{eq:cprob}.
Differentiating Relation~\eqref{eq:cprob}, we get
\be\label{eq:dotP}
\dot P_{ss'}(t)=\bra\cL(\e^{t\cL}(P_s))|P_{s'}\ket=\bra\e^{t\cL}(P_s)|\cL^\dagger(P_{s'})\ket
\ee
The simplicity of $\spec(S)$ gives that $\ran(\diag_S)=\text{span}\{P_s\mid s\in\Sigma\}$.
By Remark~\ref{remDiag}-(ii), it follows from~\DB{} that
$$
\cL^\dagger(P_{s'})=\cL^\dagger\circ\diag_S(P_{s'})=\diag_S\circ\cL^\dagger(P_{s'})=\sum_{s''}L_{s''s'}P_{s''}
$$
for some matrix $L$. Hence, Relation~\eqref{eq:dotP} becomes $\dot P(t)=P(t)L$,
and since it immediately follows from~\eqref{eq:cprob} that $P(0)=I$, we conclude that $P(t)=\e^{tL}$.
It remains to identify $L$ with the RHS of~\eqref{eq:QDef}.

Since $\bra P_{s'}|P_s\ket=\delta_{ss'}$, we deduce, using~\eqref{ldagphi},
$$
L_{s''s'}=\bra P_{s''}|\cL^\dagger(P_{s'})\ket=\bra P_{s''}|\i[H,P_{s'}]-\tfrac12\{\Phi(\un),P_{s'}\}+\Phi(P_{s'})\ket.
$$
From the proof of Lemma~\ref{comdl}, we know that $[H,P_{s'}]=[\Phi(\un),P_s]=0$,
so that, invoking the cyclicity of the trace,
$$
L_{s''s'}=\bra P_{s''}|-\Phi(\un)P_{s'}+\Phi(P_{s'})\ket
=-\delta_{s's''}\bra P_{s'}|\Phi(\un)\ket+\bra P_{s''}|\Phi(P_{s'})\ket=Q_{s''s'}.
$$
\qed

\section{2TMP Properties Inherited from Markov Processes}\label{secHeritage}

We take advantage here of the representation of outcomes of quantum measurement
processes of the observable $S^+=-\log\rho$ as a classical Markov process to get
further insight on the 2TMP distribution $\Q^t_{\rho_0}$ of that observable, in
case $\rho$ is a steady state of the Lindbladian $\cL$, under the DB condition.

\medskip

In this section we work under assumptions ensuring the validity of
Proposition~\ref{propmark}, namely: the Lindblad operator $\cL$
admits a faithful steady state $\rho$ such that the pair $(\rho,\cL)$ satisfies
the~\DB{} condition. The entropic observable $S=-\log\rho$ is assumed
to have simple spectrum $\Sigma$ and the spectral decomposition
$S=\sum_{s\in\Sigma}sP_s$.
\medskip

We will further assume the following genericity hypothesis on $S$:

\assuming{GenS}{GenS}{The numbers $s'-s$ where $s,s'\in\Sigma$ and $s\neq s'$ are all distinct.}

Under this set of hypotheses, for any faithful initial state $\rho_0$,
the 2TMP law $\Q_t^{\rho_0}$ satisfies

\begin{align*}
\Q^{t}_{\rho_0}(\Delta S=s'-s)=
\begin{cases}\bra\rho_0|P_s\ket\bra\e^{t\cL}(P_s)|P_{s'}\ket&\text{if } s'\neq s;\\[8pt]
\sum_{s\in\Sigma}\bra\rho_0|P_s\ket\bra\e^{t\cL}(P_s)|P_s\ket&\text{otherwise,}
\end{cases}
\end{align*}
since any non-vanishing variation $\Delta S$, corresponds to a unique pair $(s',s)\in\Sigma^2$,
thanks to Condition~\GenS, while a zero variation implies $s'=s$.

Hence, we have the following consequence of the classical detailed balance condition~\eqref{sympsig}: for all $s,s'\in\Sigma$
$$
\e^{-s}\Q^t_{\rho_0}(\Delta S=s'-s)\bra\rho_0|P_{s'}\ket=\e^{-s'}\Q^t_{\rho_0}(\Delta S=s-s')\bra\rho_0|P_s\ket.
$$
Equivalently, the ratio of the 2TMP probability to measure $s'-s$ and that to measure $-(s'-s)$, for $s'\neq s$, only depends
on $\e^{-(s'-s)}$ and on the ratio of the probabilities of outcomes of measures of $S$ in the initial state $\rho_0>0$,
for all $t>0$:
$$
\frac{\Q^t_{\rho_0}(\Delta S=s'-s)}{\Q^t_{\rho_0}(\Delta S=-(s'-s))}=\e^{-(s'-s)}\frac{\bra\rho_0|P_s\ket}{\bra\rho_0|P_{s'}\ket}.
$$
Choosing the initial state $\rho_0>0$ so that $\diag_S \rho_0=\rho$ we get,
for any $\sigma\neq0$,
$$
\Q^{t}_{\rho_0}(\Delta S=-\sigma)=\Q^{t}_{\rho_0}(\Delta S=\sigma),
$$
and in particular $\E^{t}_{\rho_0}(\Delta S)=0$.

\section{Moment generating function of $\Q^{t}_{\rho_0}$}\label{secMGF}

As an application of the previous Section, we consider the moment generating
function of the 2TMP distribution $\Q^{t}_{\rho_0}$ for the entropic observable
$S=-\log(\rho)$, where $\rho>0$ is a steady state of the MQD generated by $\cL$,
under the assumption~\DB{} on the pair $(\rho,\cL)$. Expressions for this moment
generating function have been derived in the literature, see {\it
e.g.}~\cite[Theorem~3.1 and Section~5]{Jaksic2014a}. The objective here is to express this moment
generating function in terms of the quantities defining the classical Markov
process attached to the 2TMP.

\begin{prop}\label{mgfL}
Let $\rho_0>0$ and assume~\DB{} holds for $(\rho,\cL)$, $S=-\log\rho$ having
simple spectrum $\Sigma$ and spectral decomposition $S=\sum_{s\in\Sigma} s P_s$.
Denoting by $d_0\in\R^{\Sigma\ast}$ the row vector
$(\bra\rho_0|P_s\ket)_{s\in\Sigma}$ and by $R$ the matrix introduced in
Remark~\ref{rem:Markov}-(iii), we have
\be\label{classgenmom}
e^t_{\rho_0}(\alpha)=\E^{t}_{\rho_0}(\e^{\alpha \Delta S})=d_0R^\alpha \e^{tQ}R^{-\alpha}{\bf 1}.
\ee
\end{prop}

\proof
Combining Relation~\eqref{eq:LalphaForm} with Lemma~\ref{comdl}, the moment generating function
can be expressed as
$$
e^t_{\rho_0}(\alpha)=\bra\e^{t\cL_\alpha}(\rho_0)|\un\ket.
$$
in terms of the deformed Lindbladian $\cL_\alpha$ defined
after~\eqref{eq:LalphaForm}. When $\dim P_s=1$, for all $s\in\Sigma$,
Formula~\eqref{classgenmom} is shown by making use of
$R^\alpha=\diag(\e^{-\alpha s})$ and Proposition~\ref{propmark} to get
\begin{align*}
e^t_{\rho_0}(\alpha)&=\sum_{s,s'\in\Sigma}\e^{\alpha (s'-s)} \bra P_s|\rho_0\ket P_{ss'}(t)\\
&=\sum_{s,s'\in\Sigma}\bra P_s|\rho_0\ket\left(R^\alpha\e^{tQ}R^{-\alpha}\right)_{ss'}
=d_0R^\alpha \e^{tQ}R^{-\alpha}{\bf 1}.
\end{align*}
\qed

Note that since $(\alpha,t)\mapsto e^t_{\rho_0}(\alpha)$ is analytic on $\C^2$, Lemma~\ref{expflu} can be rephrased as
$$
\left.\frac{\partial^2\ \ \ }{\partial t\partial\alpha}
e^t_{\rho_0}(\alpha)\right|_{(t,\alpha)=(0,0)}=\bra\rho_0|\cI\ket,
$$
where, making use of~\eqref{classgenmom}, we have
$$
\bra\rho_0|\cI\ket=-d_0Q\log R{\bf 1}.
$$
The concrete expressions for $e^t_{\rho_0}(\alpha)$ provided in Proposition~\ref{mgfL} allow for explicit computations of all moments of $\Q_{t}^{\rho_0}$, which we do not develop further. For more results about the moments generating function $e^t_{\rho_0}(\alpha)$ in case $\cL=\sum_{j\in \cJ}\cL_j$ we refer the reader to~\cite{Jaksic2014a}.

\section{Example}\label{secExample}

As an example, we consider the so called Quantum Reset Model (QRM), which is simple enough so that the quantities introduced above can be computed explicitly.
The QRM is an effective Lindbladian evolution equation arising in different guises, which is of interest in the study of so-called entanglement machines and yet simple enough to allow for a mathematical analysis. Key properties of certain QRMs are  studied in~\cite{Haack2021,Haack2024}, including entropy production. We refer to these papers for more details and consider the simplest of their versions that corresponds to the present setup.

On a $d$-dimensional Hilbert space $\cH$, the generator of the \qrm is the Lindbladian $\cL$ defined by
\be\label{renl}
\cL(\rho)=-\i[H,\rho]+\Gamma (T\tr (\rho)-\rho),
\ee
where $\Gamma>0$ and $T\in\cS$. The Hamiltonian part of the generator, $H=H^*\in\cO$, is arbitrary so far, and  we denote its repeated eigenvalues by $e_1, e_2, \dots, e_d$. For simplicity, we further make a genericity hypothesis on the spectrum
of $\ad_H(\,\cdot\;)=[H,\,\cdot\;]$ similar to~\GenS:

\assuming{Bohr}{Bohr}{The Bohr spectrum $\spec(\ad_H)\setminus\{0\}$ is simple.}

We recall some properties of the generator~\eqref{renl} of use to us, referring the reader to~\cite{Haack2021,Haack2024} for details.

\medskip
Under~\Bohr, \cite[Lemma~2.1]{Haack2021}, the spectrum of the \qrm Lindbladian is
$$
\spec(\cL)=\{0\}\cup\{-\Gamma-\i\alpha\mid\alpha\in\spec(\ad_H)\},
$$
where $0$ is simple with eigenspace spanned by
\be\label{asqrm}
\rho^+=\Gamma\big(\i\ad_H+\Gamma\big)^{-1}(T)\in\cS,
\ee
$-\Gamma$ has the $d-1$ dimensional eigenspace spanned by the traceless elements of the commutant $\{H\}'=\ker(\ad_H)$,
and the $d(d-1)$ distinct eigenvalues with non-zero imaginary parts $-\Gamma+\i\alpha$, $\alpha\in\spec(\ad_H)\setminus\{0\}$,
appear as complex conjugate pairs. Moreover,  one has the explicit expression,
\be\label{basisindep}
\e^{t\cL}(\rho_0)=\bra\rho_0|\un\ket\rho^+
+\e^{-t\Gamma}\e^{-\i tH}\big(\rho_0-\bra\rho_0|\un\ket\rho^+ \big)\e^{\i tH},
\ee
for any $\rho_0\in\cT$. In particular, the MQD generated by $\cL$ is relaxing with asymptotic state $\rho^+$.
It is a simple matter to check that the CP map associated to $\cL$ is given by
$$
\Phi(X)=\Gamma\bra T|X\ket\un.
$$
The following properties are proven in~\cite[Lemmas~3.1 and 3.3]{Haack2024}:

\begin{lem}
For any $H=H^*$, and $T\in\cT$, the linear map  $\rho_H:\cT\to\cT$
$$
T\mapsto\rho_H(T)=\big(\i\ad_H+\Gamma\big)^{-1}(T),$$ see~\eqref{asqrm}, is CPTP
and such that $T>0\Longrightarrow\rho_H(T)>0$.

For $T>0$, the Lindbladian $\cL$ defined by~\eqref{renl} and its asymptotic state $\rho^+$ given by~\eqref{asqrm} it holds:
\begin{enumerate}[label=(\roman*)]
\item The pair $(\rho^+,\cL)$ satisfies the detailed balance condition~\DB{} if and only if $[H,T]=0$.\\
\item If condition~\DB{} holds, $\rho^+=T$ and the EP~\eqref{singlep} of the MQD generated by $\cL$ in the state $\rho$
reads
$$
\EP(\rho)=\Gamma (\Ent(T|\rho)+\Ent(\rho |T))
$$
for all faithful $\rho\in\cS$, so that $\EP(\rho)=0$ if and only if $\rho=T$.
\end{enumerate}
\end{lem}

Hence we assume from now on that $T>0$ and that~\DB{} holds, so that for all $\rho_0\in\cS$,
$$
\e^{t\cL}(\rho_0)=\e^{-\Gamma t}\e^{-\i tH}\rho_0\e^{\i tH} + T(1-\e^{-\Gamma t}),
$$
which shows that $\e^{t\cL}$ is positivity improving. In particular, under these assumptions,
the entropic observable of the QRM model is $S=-\log T$, and for any $\rho_0=\diag_S (\rho_0)\in\cS$
we have $[H,\rho_0]=0$, so that
\be\label{evqrmdb}
\e^{t\cL}(\diag_S(\rho_0))=T+\e^{-t\Gamma}(\diag_S(\rho_0)-T).
\ee
In particular, the matrix elements of $P(t)$ read with $S=\sum_{s}sP_s$
$$
P_{ss'}(t)=\bra\e^{t\cL}(P_s)|P_{s'}\ket=\e^{-s'}(1-\e^{-t\Gamma})+\e^{-t\Gamma}\delta_{ss'}.
$$
This yieds the spectral decomposition of $P(t)$  in term of the vectors ${\bf 1}$ and $\pi^+=(\e^{-s})$ s.t.\;$\pi^+{\bf 1}=1$, $({\bf 1}\pi^+)_{ss'}=\e^{-s'}$,
$$
P(t)=\e^{-t\Gamma}(\un-{\bf 1}\pi^+ )+{\bf 1} \pi^+ ,
$$
and therefore
$$
Q=-\Gamma(\un-{\bf 1} \pi^+ ),\quad\text{ so that }\quad Q_{ss'}=\Gamma (\e^{-s'}-\delta_{ss'}).
$$
In turn, we obtain the sought for explicit 2TMP distribution $\Q^t_{\rho_0}$
\begin{prop}\label{prop:jason}
Under the hypotheses of Section~\ref{secHeritage}, in particular~\GenS, we have for the
QRM model~\eqref{renl}, for any state $\rho_0>0$ and $S=-\log T$,
$$
\Q^t_{\rho_0}(\Delta S=\sigma)=\begin{cases}
(1-\e^{-t\Gamma})\bra\rho_0|P_s\ket\e^{-s'}&\text{if }\sigma=s'-s\neq0;\\[8pt]
(1-\e^{-t\Gamma})\bra\rho_0|T\ket+\e^{-t\Gamma}&\text{otherwise}.
\end{cases}
$$
\end{prop}
\begin{rem}
\begin{enumerate}[label=(\roman*)]
\item Without assuming~\GenS, one has
\begin{align*}
\E^{t}_{\rho_0}(\Delta S)&=(1-\e^{-t\Gamma })\bra T-\diag_S(\rho_0)|S\ket,\\
e^t_{\rho_0}(\alpha)&=\bra\rho_0|T^\alpha\ket\bra T^{1-\alpha}|\un\ket(1-\e^{-t\Gamma})+\e^{-t\Gamma}.
\end{align*}
The first formula is a straightforward consequence of~\eqref{exp2TDB} and~\eqref{evqrmdb}, whereas the second one stems from~\eqref{basisindep}.
\item In case $\diag_S(\rho_0)=T$, we recover from the previous formula that $\E_{\Q_{t}^{\rho_0}}(\Delta S)=0$, in accordance with~\eqref{exp2TDB}.
\end{enumerate}
\end{rem}

Finally, to tackle the many-reservoir setup characterized by different dissipators,
within the context of \qrm, we consider the following construction, see~\cite[Section~4]{Haack2024}.

Let $\cJ$ be a finite set of indices, for  $j\in\cJ$ let $\lambda_j\in\R$ be such that
$\sum_{j\in\cJ}\lambda_j=1$, and set
$$
\cL_j(\rho)=-\i[\lambda_jH,\rho]+\Gamma_j(T_j \,\tr(\rho)-\rho),
$$
where  the dissipator is characterized by $T_j\in\cS$ and the coupling rates $\Gamma_j>0$. Then,
the full Lindbladian $\cL=\sum_{j\in\cJ}\cL_j$ is given by
$$
\cL (\rho)=-\i[H,\rho]+\Gamma (T\,\tr(\rho)-\rho)
$$
where
$$
\Gamma=\sum_{j\in\cJ}\Gamma_j>0,\qquad T=\frac{1}{\Gamma}\sum_{j\in\cJ}\Gamma_jT_j\in\cS,
$$
and reduces to~\eqref{renl}. Thus, assuming $[H,T_j]=0$ for all $j\in\cJ$, \DB{} holds for
$(T_j,\cL_j)$ and $(T,\cL)$.

Thus, Theorem~\ref{ept=0} applies to yield with $S_j^+=-\log(T_j)$ and Proposition~\ref{prop:jason}
$$
\EP(T)=-\sum_{j\in\cJ}\left.\frac{\d\ }{\d t}\E_{j,T}^t(\Delta S_j^+)\right|_{t=0}
=\sum_{j\in\cJ}\Gamma_j(\Ent(T|T_j)+\Ent(T_j|T)).
$$
Similarly, by Relation~\eqref{epvs2tmp}
$$
\sum_{j\in\cJ}\int_0^\infty\EP_j(\e^{s\cL_j}(T))\d s=-\sum_{j\in\cJ}\E_{\Q_{\infty,j}^{T}}(\Delta S_j^+)
+\sum_{j\in\cJ}(S(T_j)-S(T))=\sum_{j\in\cJ}\Ent(T|T_j).
$$

\section{Proof of Proposition~\ref{prop:hebo}}
\label{sec:proof5.1}

Assume that the MQD generated by $\cL$ is relaxing, the initial
state $\rho$ and the asymptotic state $\rho^+$ being both faithful. In our
finite dimensional context, the map $\C\ni t\mapsto\rho(t)=\e^{t\cL}(\rho)$ is
entire, taking its values in the self-adjoint elements of $\cS$ for $t\ge0$. In
particular, the spectral decomposition
$$
\rho(t)=\sum_{h=1}^m p_h(t)P_h(t),
$$
is such that the eigenvalues $p_h(t)$ and eigenprojections $P_h(t)$ are real-analytic
functions of $t$, even at exceptional points where some eigenvalues
coincide~\cite[Chapter~2, Theorem~1.10]{Kato1966}.

Expressing the von Neumann entropy
in terms of the function $[0,1]\ni p\mapsto\chi(p)=-p\log(p)$,\footnote{Note
that the multiplicities $g_h$ are (integer) constants.}
\be\label{mvne}
S(\rho(t))=\sum_{h=1}^m g_h \chi(p_h(t)),\qquad g_h=\tr P_h(t),
\ee
with the usual convention that $\chi(0)=0$ and $\chi'(0^+)=+\infty$, we
have that $S(\rho(t))$ is finite for all $t\geq 0$, but its time-derivative might
diverge whenever some eigenvalues $p_h(t)$ vanish. We show that this does
not happen.

First note that for small enough $t\ge0$, $\rho(t)>0$ by continuity, whereas the
relaxing assumption ensures that $\rho(t)>0$ for large enough $t$. Thus, we can
restrict our attention to $t\in(-\delta, 1/\delta)$, for some $\delta>0$ small enough.
Let $t_0\in (-\delta, 1/\delta)$ be such that some eigenvalue
$p(t)$ vanishes at $t=t_0$, (dropping the index from the notation). We can
focus on the corresponding contribution of that eigenvalue to~\eqref{mvne}.
The non-negativity of $\rho(t)$ implies that in a neighborhood of $t_0$
the analytic function $p(t)$ factorizes as
\be
p(t_0+\tau)=\tau^{2p}r(\tau),
\ee
where $p\in\N^\ast$ and the function $r$ is analytic near $0$ and such that $a=r(0)>0$.
It follows that
$$
\chi(p(t_0+\tau))=-ap\tau^{2p}(\log\tau^2+O(1)),\quad(\tau\to0),
$$
so that
\be
\left.\frac{\d\ }{\d t}\chi(p(t))\right|_{t=t_0}
=-\lim_{\tau\to0}ap\tau^{2p-1}(\log\tau^2+O(1))=0.
\ee
Altogether, we showed that the function $t\mapsto S(\rho(t))$ is of class $C^1$ on $(0,\infty)$.
Since $\rho(0)>0$ by assumption, it has a finite right derivative at $t=0$. This proves
part~(i). Part~(ii) follows immediately since $\rho(t)>0$ for large enough $t$.

\bigskip
\noindent {\bf Acknowledgments.} This work is partially supported by the French National Research Agency in the framework of the
"France 2030" program (ANR-11-LABX-0025-01) for the LabEx PERSYVAL, and by CY Initiative, Investissements d’Avenir, grant number ANR-16-IDEX-0008.

\medskip
\noindent {\bf Conflict of Interest Statement.} 
On behalf of all authors, the corresponding author states that there is no conflict of interest.

\bibliographystyle{capalpha}
\bibliography{MASTER}

\newcommand{\etalchar}[1]{$^{#1}$}
\providecommand{\bysame}{\leavevmode \hbox to3em{\hrulefill}\thinspace}
\providecommand{\og}{``}
\providecommand{\fg}{''}
\providecommand{\smfandname}{and}
\providecommand{\smfedsname}{eds.}
\providecommand{\smfedname}{ed.}
\providecommand{\smfmastersthesisname}{Master Thesis}
\providecommand{\smfphdthesisname}{Thesis}
\begin{thebibliography}{BBJ{\etalchar{+}}24b}

\bibitem[Aga73]{Agarwal1973}
{\scshape Agarwal, G.~S.}: Open quantum {M}arkovian systems and the
  microreversibility. Z. Physik \textbf{258}, 409--422 (1973),
  \href{https://doi.org/10.1007/bf01391504}{[DOI:10.1007/bf01391504]}.

\bibitem[Ali76]{Alicki1976}
{\scshape Alicki, R.}: On the detailed balance condition for non-{H}amiltonian
  systems. Rep. Math. Phys. \textbf{10}, 249--258 (1976),
  \href{https://doi.org/10.1016/0034-4877(76)90046-X}{[DOI:10.1016/0034-4877(76)90046-X]}.

\bibitem[BBJ{\etalchar{+}}23]{Benoist2023a}
{\scshape Benoist, T., Bruneau, L., {Jak\v si\'c}, V., Panati, A. {\normalfont
  \smfandname} Pillet, C.-A.}: A note on two-times measurement entropy
  production and modular theory. Lett. Math. Phys. \textbf{114:32}, (2023),
  \href{https://doi.org/10.1007/s11005-024-01777-0}{[DOI:10.1007/s11005-024-01777-0]}.

\bibitem[BBJ{\etalchar{+}}24a]{Benoist2024}
{\scshape Benoist, T., Bruneau, L., {Jak\v si\'c}, V., Panati, A. {\normalfont
  \smfandname} Pillet, C.-A.}: Entropic fluctuations in statistical mechanics
  {II}. {Q}uantum dynamical systems. Preprint, 2024,
  \href{https://doi.org/10.48550/arXiv.2409.15485}{[DOI:10.48550/arXiv.2409.15485]}.

\bibitem[BBJ{\etalchar{+}}24b]{Benoist2024b}
\bysame : On the thermodynamic limit of two-times measurement entropy
  production. Preprint, 2024,
  \href{https://doi.org/10.48550/arXiv.2402.09380}{[DOI:10.48550/arXiv.2402.09380]}.

\bibitem[BHR22]{Benoist2022}
{\scshape Benoist, T., Hänggli, L. {\normalfont \smfandname} Rouzé, C.}:
  Deviation bounds and concentration inequalities for quantum noises. Quantum
  \textbf{6}, 772 (2022),
  \href{https://doi.org/10.22331/q-2022-08-04-772}{[DOI:10.22331/q-2022-08-04-772]}.

\bibitem[CM17]{Carlen2017}
{\scshape Carlen, E.~A. {\normalfont \smfandname} Maas, J.}: Gradient flow and
  entropy inequalities for quantum markov semigroups with detailed balance. J.
  Funct. Anal. \textbf{273}, 1810--1869 (2017),
  \href{https://doi.org/10.1016/j.jfa.2017.05.003}{[DOI:10.1016/j.jfa.2017.05.003]}.

\bibitem[Dav74]{Davies1974}
{\scshape Davies, E.~B.}: Markovian master equations. Commun. Math. Phys.
  \textbf{39}, 91--110 (1974),
  \href{https://doi.org/10.1007/bf01608389}{[DOI:10.1007/bf01608389]}.

\bibitem[Dav76]{Davies1976a}
\bysame : Markovian master equations. {II}. Math. Ann. \textbf{219}, 147--158
  (1976), \href{https://doi.org/10.1007/BF01351898}{[DOI:10.1007/BF01351898]}.

\bibitem[DDRM08]{Derezinski2008}
{\scshape Derezi\'{n}ski, J., De~Roeck, W. {\normalfont \smfandname} Maes, C.}:
  Fluctuations of quantum currents and unravelings of master equations. J.
  Stat. Phys. \textbf{131}, 341--356 (2008),
  \href{https://doi.org/10.1007/s10955-008-9500-8}{[DOI:10.1007/s10955-008-9500-8]}.

\bibitem[FGM23]{Fiorelli2023}
{\scshape Fiorelli, E., Gherardini, S. {\normalfont \smfandname} Marcantoni,
  S.}: Stochastic entropy production: Fluctuation relation and irreversibility
  mitigation in non-unital quantum dynamics. Journal of Statistical Physics
  \textbf{190}, (2023),
  \href{https://doi.org/10.1007/s10955-023-03118-2}{[DOI:10.1007/s10955-023-03118-2]}.

\bibitem[FU07]{Fagnola2007}
{\scshape Fagnola, F. {\normalfont \smfandname} Umanità, V.}: Generators of
  detailed balance quantum {M}arkov semigroups. Infinite Dimensional Analysis,
  Quantum Probability and Related Topics \textbf{10}, 335--363 (2007),
  \href{https://doi.org/10.1142/S0219025707002762}{[DOI:10.1142/S0219025707002762]}.

\bibitem[FU10]{Fagnola2010}
{\scshape Fagnola, F. {\normalfont \smfandname} Umanit{\`{a}}, V.}: Generators
  of {KMS} symmetric {M}arkov semigroups on {$\mathcal{B}(\mathrm{h})$}.
  {S}ymmetry and quantum detailed balance. Commun. Math. Phys. \textbf{298},
  523--547 (2010),
  \href{https://doi.org/10.1007/s00220-010-1011-1}{[DOI:10.1007/s00220-010-1011-1]}.

\bibitem[GKS76]{Gorini1976}
{\scshape Gorini, V., Kossakowski, A. {\normalfont \smfandname} Sudarshan, E.
  C.~G.}: Completely positive dynamical semigroups of {$N$}-level systems. J.
  Math. Phys. \textbf{17}, 821--825 (1976),
  \href{https://doi.org/10.1063/1.522979}{[DOI:10.1063/1.522979]}.

\bibitem[HJ21]{Haack2021}
{\scshape Haack, G. {\normalfont \smfandname} Joye, A.}: Perturbation analysis
  of quantum reset models. Journal of Statistical Physics \textbf{183}, (2021),
  \href{https://doi.org/10.1007/s10955-021-02752-y}{[DOI:10.1007/s10955-021-02752-y]}.

\bibitem[HJ24]{Haack2024}
\bysame : Entropy production of quantum reset models. Preprint, 2024,
  \href{https://doi.org/10.48550/arXiv.2401.10022}{[DOI:10.48550/arXiv.2401.10022]}.

\bibitem[JOPP10]{Jaksic2010b}
{\scshape Jak{\v{s}}i{\'{c}}, V., Ogata, Y., Pautrat, Y. {\normalfont
  \smfandname} Pillet, C.-A.}: {Entropic fluctuations in quantum statistical
  mechanics---{A}n introduction}. In \emph{{Quantum Theory from Small to Large
  Scales}} (Fr{\"{o}}hlich, J., Salmhofer, M., de~Roeck, W., Mastropietro, V.
  {\normalfont \smfandname} Cugliandolo, L., \smfedsname), {Lecture Notes of
  the Les Houches Summer School}, vol.~95, Oxford University Press, Oxford,
  p.~213--410, 2010,
  \href{https://doi.org/10.1093/acprof:oso/9780199652495.003.0004}{[DOI:10.1093/acprof:oso/9780199652495.003.0004]}.

\bibitem[JPW14]{Jaksic2014a}
{\scshape Jak\v{s}i\'{c}, V., Pillet, C.-A. {\normalfont \smfandname} Westrich,
  M.}: Entropic fluctuations of quantum dynamical semigroups. J. Stat. Phys.
  \textbf{154}, 153--187 (2014),
  \href{https://doi.org/10.1007/s10955-013-0826-5}{[DOI:10.1007/s10955-013-0826-5]}.

\bibitem[Kat66]{Kato1966}
{\scshape Kato, T.}: \emph{Perturbation theory for linear operators}. Die
  Grundlehren der mathematischen Wissenschaften, Band 132, Springer, New York,
  1966,
  \href{https://doi.org/10.1007/978-3-662-12678-3}{[DOI:10.1007/978-3-662-12678-3]}.

\bibitem[KFGV77]{Kossakowski1977}
{\scshape Kossakowski, A., Frigerio, A., Gorini, V. {\normalfont \smfandname}
  Verri, M.}: Quantum detailed balance and {KMS} condition. Commun. Math. Phys.
  \textbf{57}, 97--110 (1977),
  \href{https://doi.org/10.1007/bf01625769}{[DOI:10.1007/bf01625769]}.

\bibitem[Kur00]{Kurchan2000}
{\scshape Kurchan, J.}: A quantum fluctuation theorem. Unpublished, 2000,
  \href{https://doi.org/10.48550/arXiv.cond-mat/0007360}{[DOI:10.48550/arXiv.cond-mat/0007360]}.

\bibitem[Lin76]{Lindblad1976}
{\scshape Lindblad, G.}: On the generators of quantum dynamical semigroups.
  Commun. Math. Phys. \textbf{48}, 119--130 (1976),
  \href{https://doi.org/10.1007/bf01608499}{[DOI:10.1007/bf01608499]}.

\bibitem[LLL96]{Levitov1996}
{\scshape Levitov, L.~S., Lee, H. {\normalfont \smfandname} Lesovik, G.~B.}:
  Electron counting statistics and coherent states of electric current. J.
  Math. Phys. \textbf{37}, 4845--4866 (1996),
  \href{https://doi.org/10.1063/1.531672}{[DOI:10.1063/1.531672]}.

\bibitem[Nor97]{Norris1997}
{\scshape Norris, J.~R.}: \emph{Markov chains}. Cambridge University Press,
  1997,
  \href{https://doi.org/10.1017/cbo9780511810633}{[DOI:10.1017/cbo9780511810633]}.

\bibitem[SL78]{Spohn1978b}
{\scshape Spohn, H. {\normalfont \smfandname} Lebowitz, J.~L.}: Irreversible
  thermodynamics for quantum systems weakly coupled to thermal reservoirs. Adv.
  Chem. Phys. \textbf{38}, 109--142 (1978),
  \href{https://doi.org/10.1002/9780470142578.ch2}{[DOI:10.1002/9780470142578.ch2]}.

\bibitem[Spo77]{Spohn77b}
{\scshape Spohn, H.}: An algebraic condition for the approach to equilibrium of
  an open {$N$}-level system. Lett. Math. Phys. \textbf{2}, 33--38 (1977),
  \href{https://doi.org/10.1007/BF00420668}{[DOI:10.1007/BF00420668]}.

\bibitem[Spo78]{Spohn1978a}
\bysame : Entropy production for quantum dynamical semigroups. J. Math. Phys.
  \textbf{19}, 1227--1230 (1978),
  \href{https://doi.org/10.1063/1.523789}{[DOI:10.1063/1.523789]}.

\end{thebibliography}

\end{document}